\pdfoutput=1
\documentclass[prb,twocolumn,showpacs,amsmath,amssymb,floatfix]{revtex4}

\usepackage{graphicx}%Include figure files
\usepackage{dcolumn}%Align table columns on decimal point
\usepackage{bm}% bold math
\begin{document}

\title{Edge currents and nanopore arrays in zigzag and chiral graphene nanoribbons as a route toward high-$ZT$ thermoelectrics}

\author{Po-Hao Chang}
\author{Branislav K. Nikoli\' c}
\affiliation{Department of Physics and Astronomy, University of Delaware, Newark, DE 19716-2570, USA}

\begin{abstract}
We analyze electronic and phononic quantum transport through zigzag or chiral graphene nanoribbons (GNRs) perforated with an array of nanopores. Since local charge current profiles in these GNRs are peaked around their edges, drilling nanopores in their interior does not affect such edge charge currents while drastically reducing heat current carried by phonons in sufficiently long wires. The combination of these two effects can yield highly efficient thermoelectric devices with maximum \mbox{$ZT \simeq 11$} at liquid nitrogen temperature and \mbox{$ZT \simeq 4$} at room temperature achieved in \mbox{$\sim 1$ $\mu$m} long zigzag GNRs with nanopores of variable diameter and spacing between them. Our analysis is based on the $\pi$-orbital tight-binding Hamiltonian with up to third nearest-neighbor hopping for electronic subsystem, the empirical fourth-nearest-neighbor model for phononic subsystem, and nonequilibrium Green function formalism to study quantum transport in both of these models.
\end{abstract}

\maketitle

The recent explosion of research on graphene---one-atom-thick allotrope of carbon---has been largely focused on its unique electronic structure and transport properties governed by the two-dimensional honeycomb lattice of carbon atoms.~\cite{Geim2009} Very recently, the exploration of its thermal and thermoelectric properties has been initiated by measuring  the thermopower~\cite{Zuev2009,Wei2009,Checkelsky2009} $S$ and phonon thermal conductivity~\cite{Balandin2011} $K_{\rm ph}$ of large-area graphene. The measured values~\cite{Zuev2009} of $S \simeq 100$ $\mu$V/K near the Dirac point (DP), as well as the room-temperature $K_{\rm ph} \simeq 4000$ W/mK (averaged over values obtained using different samples and experimental techniques~\cite{Balandin2011}) which outperforms virtually all other known materials, point out that {\em large-area} graphene is {\em not suitable} for thermoelectric applications.

Thermoelectrics transform temperature gradients into electric voltage and vice versa. Although a plethora of widespread applications has been envisioned, their usage is presently limited by their small efficiency.~\cite{Vining2009} Thus, careful tradeoffs are required to optimize the dimensionless figure of merit
\begin{equation}\label{eq:zt}
ZT=\frac{S^2GT}{\kappa_{\rm el} + \kappa_{\rm ph}},
\end{equation}
which quantifies the maximum efficiency of a thermoelectric cycle conversion in the linear-response regime where small voltage $V=-S \Delta T$  exactly cancels the current induced by the small thermal bias $\Delta T$. This is due to the fact that $ZT$ contains unfavorable combination of $S$, average temperature $T$, electronic conductance $G$ and thermal conductance $\kappa_{\rm el} + \kappa_{\rm ph}$. The total thermal conductance has contributions from both electrons $\kappa_{\rm el}$ and phonons $\kappa_{\rm ph}$. The devices with $ZT > 1$ are regarded as good thermoelectrics, but values of $ZT > 3$  are required for thermoelectric devices to compete in efficiency with conventional power generators and refrigerators.~\cite{Vining2009}

Thus, a number of proposals have been put forth to evade the problem of high lattice thermal conductivity of large-area graphene that could open a pathway for its thermoelectric applications. For example, large-area graphene could reach $ZT \approx 0.3$ if perforated by the so-called antidot lattice tailored to impede phonon propagation.~\cite{Gunst2011} Switching to quasi-one-dimensional graphene nanoribbons (GNRs) makes possible further enhancement of $ZT$ where it has been predicted that long ($\sim 1$ $\mu$m) GNRs with zigzag edges and disorder introduced along such edges could reach $ZT \simeq 4$ at room temperature.~\cite{Sevincli2010} Another route is to engineer structural defects in GNRs that can block phonons while retaining quasiballistic electronic transport.~\cite{Haskins2011}

However, it is more advantageous to search for high-$ZT$ devices among clean nanowires~\cite{Markussen2009} since edge or surface disorder can affect electronic conductance significantly. For example,  the experiments on  etched GNRs with rough edges find  Coulomb blockade effects (not taken into account in Ref.~\onlinecite{Sevincli2010}) and transport gap much larger than the band gap.~\cite{Molitor2010}

\begin{figure}
\includegraphics[scale=0.5,angle=0]{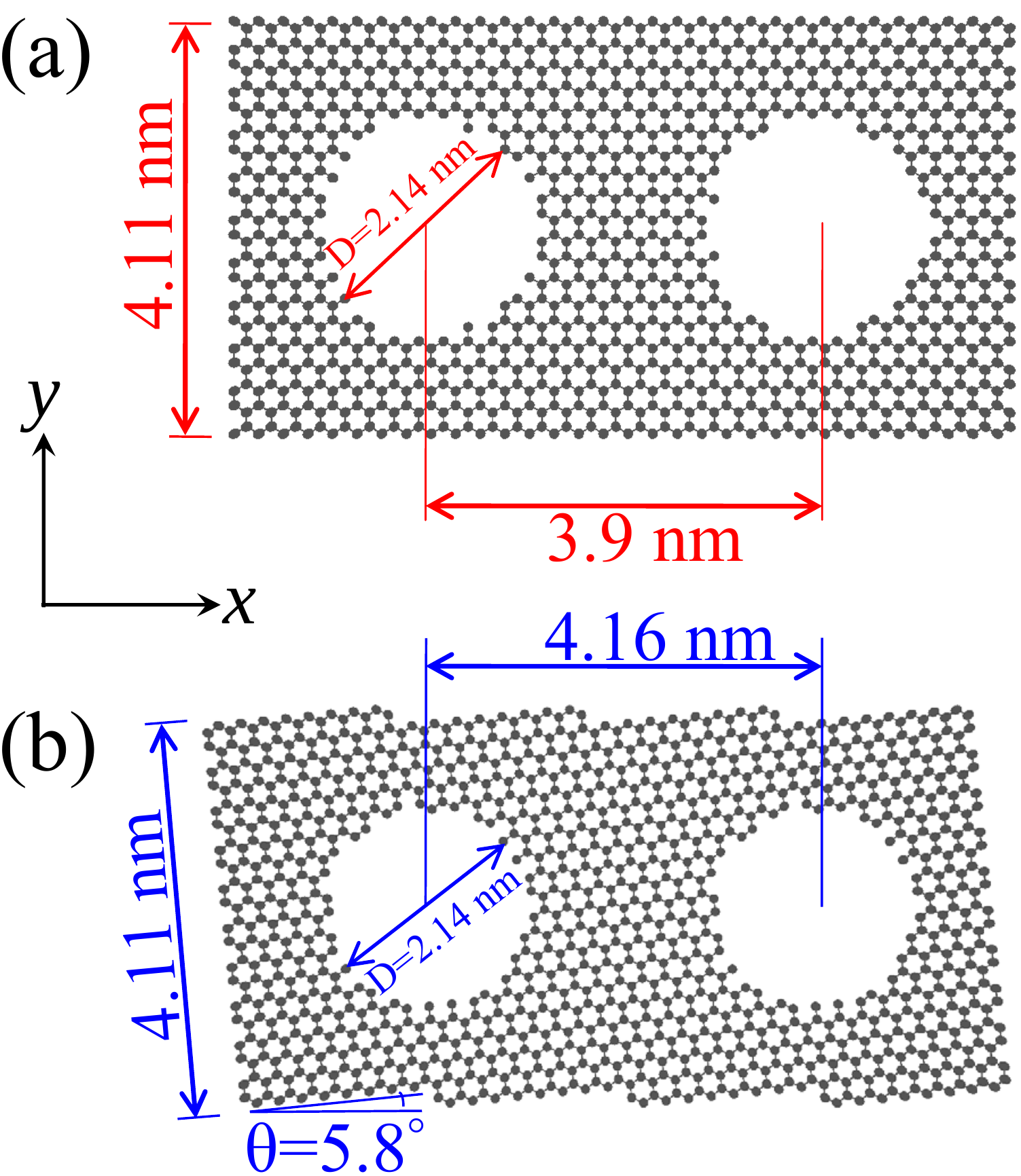}
\caption{(Color online) Schematic view of: (a) 20-ZGNR (composed of 20 zigzag chains); and (b) (8,1)-CGNR with chiral angle $\theta=5.8^\circ$. The size of the nanopores, assumed to be drilled in the GNR interior away from its zigzag or chiral edges, and the distance between them is illustrated by plotting two repeated supercells of each GNR. The length of these GNRs in actual calculations is set to $L \simeq 1.2$ $\mu$m.}
\label{fig:fig1}
\end{figure}

\begin{figure}
\includegraphics[scale=0.3,angle=0]{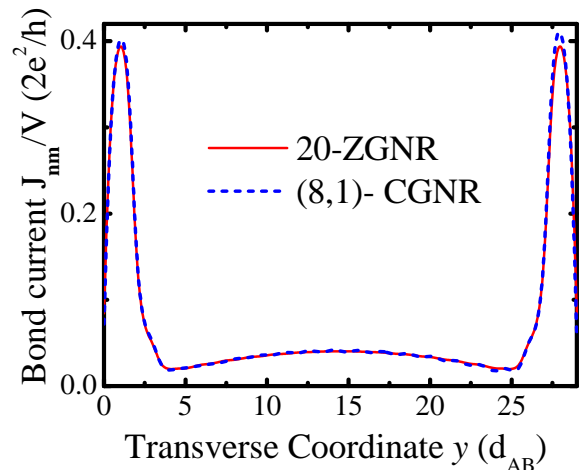}
\caption{(Color online) Spatial profile of local charge currents over the transverse cross section of infinite homogeneous 20-ZGNR or (8,1)-CGNR for electronic transport close ($E_F=-0.43$ eV) to the DP. The sum of bond currents~\cite{Zarbo2007} $J_{\bf nm}/V$, which describe charge flow from site ${\bf n}$ to site ${\bf m}$ of the honeycomb lattice if  hopping $t_{\bf n}^{\bf m} \neq 0$ is non-zero between the two sites, gives the conductance $G=I/V$ ($I$ is the total current in the leads and  $V \rightarrow 0$ is small bias voltage driving the linear-response transport).}
\label{fig:fig2}
\end{figure}

\begin{figure}
\includegraphics[scale=0.33,angle=0]{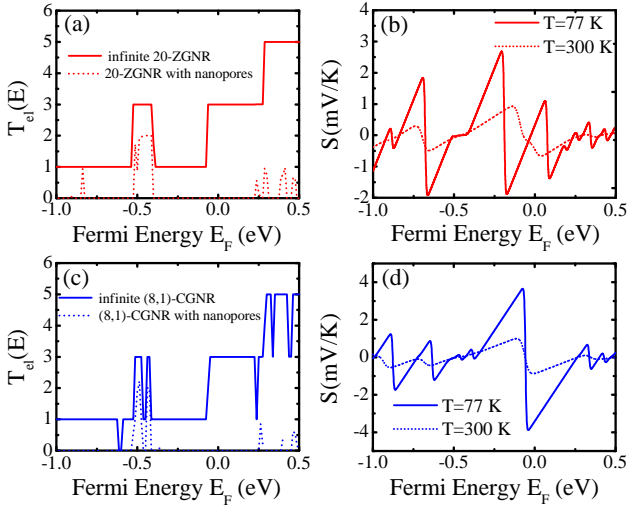}
\caption{(Color online) (a) Zero-bias electronic transmission $\mathcal{T}_{\rm el}(E)$ for an infinite homogeneous 20-ZGNR or 20-ZGNR of length $L \simeq 1.2$ $\mu$m with periodic array of identical nanopores shown in Fig.~\ref{fig:fig1}(a). (b) Thermopower at two different temperatures for finite length 20-ZGNR with nanopores. (c) Zero-bias electronic transmission for an infinite homogeneous  (8,1)-CGNR or (8,1)-CGNR of length $L \simeq 1.2$ $\mu$m with periodic array of identical nanopores shown in Fig.~\ref{fig:fig1}(b). (d) Thermopower at two different temperatures for finite length (8,1)-CGNR with nanopores.}
\label{fig:fig3}
\end{figure}

In this Rapid Communication, we exploit peculiar electronic transport properties of {\em clean} GNRs with zigzag (ZGNR) or chiral (CGNR) edges,  illustrated in Fig.~\ref{fig:fig1}, where the local charge current density carried by quasiparticles sufficiently close to the DP is peaked around nanoribbons edges as demonstrated in Fig.~\ref{fig:fig2}. Thus, drilling nanopores~\cite{Tada2011} in the ZGNR or CGNR interior will not substantially modify such ``edge currents''. This is confirmed by the transmission function in Figs.~\ref{fig:fig3}(a) and ~\ref{fig:fig3}(c) which is reduced from $\mathcal{T}_{\rm el}(E)=3$ in homogeneous GNRs to $\mathcal{T}_{\rm el}(E) \simeq 2$ around the DP for both ZGNR and CGNR with an array of nanopores. Furthermore, $\mathcal{T}_{\rm el}(E)$ around the DP does not change as one increases the length of GNRs because ``edge currents'' propagate quasiballistically.

\begin{figure}
\includegraphics[scale=0.33,angle=0]{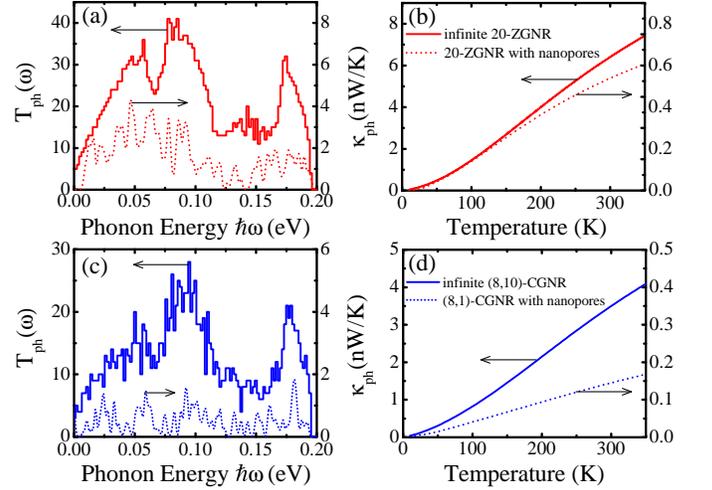}
\caption{(Color online) (a) The phonon transmission function $\mathcal{T}_{\rm ph}(\omega)$ and (b) the corresponding phonon thermal conductance $\kappa_{\rm ph}$ for an infinite 20-ZGNR or 20-ZGNR of length $L \simeq 1.2$ $\mu$m with periodic array of identical nanopores shown in Fig.~\ref{fig:fig1}(a). (c) The phonon transmission function and (d) the corresponding phonon thermal conductance $\kappa_{\rm ph}$ for an infinite (8,1)-CGNR or (8,1)-ZGNR of length $L \simeq 1.2$ $\mu$m with a periodic array of identical nanopores shown in Fig.~\ref{fig:fig1}(b).}
\label{fig:fig4}
\end{figure}

On the other hand, nanopore arrays, whose fabrication has been pursued recently by a variety of experimental techniques,~\cite{Tada2011} break  homogeneity of the nanowire so that they can substantially impede the propagation of phonons in sufficiently long GNRs. This is corroborated by our results for the phonon transmission function in Figs.~\ref{fig:fig4}(a) and \ref{fig:fig4}(b) and the corresponding lattice thermal conductance in Figs.~\ref{fig:fig4}(c) and ~\ref{fig:fig4}(d).

\begin{figure}
\includegraphics[scale=0.33,angle=0]{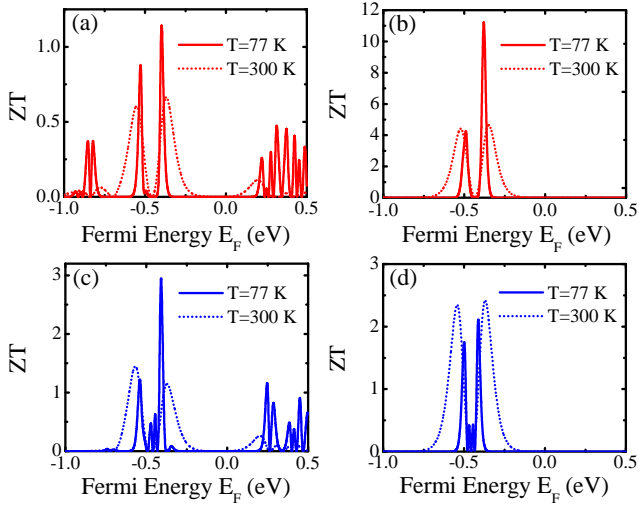}
\caption{(Color online) The thermoelectric figure of merit $ZT$ for GNR+nanopores devices of length $L \simeq 1.2$ $\mu$m: (a) $ZT$ vs. energy at two different temperatures for 20-ZGNR with a periodic array of  identical nanopores illustrated in Fig.~\ref{fig:fig1}(a); (b) $ZT$ vs. energy at  two different temperatures for 20-ZGNR with pores of variable diameter $D \in [4.5 \, d_{AB},  7.5 \, d_{AB}]$ and their position shifted by a value $\Delta x \in [-2 \, d_{AB}, 2 \, d_{AB}]$ along the transport direction away from the original position within periodic array shown in Fig.~\ref{fig:fig1}(a); (c) $ZT$ vs. energy at two different temperatures for (8,1)-CGNR with a periodic array of  identical nanopores illustrated in Fig.~\ref{fig:fig1}(b); and (d) $ZT$ vs. energy at  two different temperatures for (8,1)-CGNR with pores of variable diameter $D \in [1.5 \,  d_{AB},  3.5 \, d_{AB}]$ and their position shifted by a value $\Delta x \in [-0.5 \, d_{AB}, 0.5 \, d_{AB}]$ along the transport direction away from the original position within periodic array shown in Fig.~\ref{fig:fig1}(b).}
\label{fig:fig5}
\end{figure}

Combining these two effects, we obtain maximum $ZT \simeq 3$ at $T=77$ K and  $ZT \simeq 1.5$ at $T=300$ K in Fig.~\ref{fig:fig5}(c) for the case of (8,1)-CGNR in Fig.~\ref{fig:fig1} whose identical nanopores are arranged in a perfectly symmetric and ordered array. The values of $ZT$ for 20-ZGNR with periodic array of nanopores are lower, as shown in Fig.~\ref{fig:fig5}(a). In realistic GNR-based devices, it may be challenging~\cite{Tada2011} to control the pore arrangement to a high degree of order shown in Fig.~\ref{fig:fig1}. If we assume that pore diameter takes a random value within some interval and that nanopores are not arranged into perfectly periodic array, then we find in Fig.~\ref{fig:fig5}(b) a possibility of enormous $ZT \simeq 11$  at $T=77$ K and $ZT \simeq 4$ at $T=300$ K in the case of ZGNR-based device.

In the rest of the paper we explain details of our models for electronic and phononic subsystem and the corresponding quantum transport calculations. The early theoretical studies of ZGNR-based devices have utilized~\cite{Rycerz2007}  a simplistic tight-binding model (TBM) with single $\pi$-orbital per site and the nearest-neighbor hopping only, or its long-wavelength (continuum) approximation---the  Dirac-Weyl Hamiltonian~\cite{Brey2006}---valid close to the DP. However, both of these models predict~\cite{Zarbo2007,Rycerz2007}  that transmission function of an infinite homogeneous ZGNR is \mbox{$\mathcal{T}_{\rm el}=1$} around the DP and that current density profile is peaked~\cite{Zarbo2007} in the middle of ZGNR (even though local density of states reaches maximum around the edges). This contradicts first-principles calculations~\cite{Saha2011a}, or TBM with up to third nearest-neighbor~\cite{Cresti2008} hopping parameters fitted to such first-principles calculations, which predict $\mathcal{T}_{\rm el}=3$ around the DP, as well as that local current density is mostly confined to flow around the zigzag edges.~\cite{Areshkin2007a} It is worth mentioning that the majority of recent studies of thermoelectric properties of ZGNRs with edge disorder~\cite{Sevincli2010} or of finite length graphene antidot lattice~\cite{Gunst2011} have utilized the TBM with nearest-neighbor hopping, so that a possibility to exploit ``edge currents'' around zigzag or chiral edges for thermoelectric device applications has been overlooked.

Most importantly, the recent experiments have confirmed the existence of ``edge currents'' in metallic ZGNRs by actually utilizing them  to increase the heat dissipation around edge defects and, thereby, rearrange atomic structure locally until sharply defined zigzag edge is achieved.~\cite{Jia2009} Also, the very recent chemical synthesis~\cite{Tao2011} of (8,1)-CGNRs via carbon nanotube unzipping method have exhibited  properties in sub-nanometer-resolved scanning tunneling microscopy and spectroscopy that can only be explained by the existence of smooth edges supporting edge quantum states (i.e., wavefunctions whose probability density is large around the edges). Although ZGNRs or CGNRs~\cite{Tao2011} are insulating at very low temperatures due to one-dimensional spin-polarized edge states coupled across the width of the nanoribbon, such unusual magnetic ordering and the corresponding band gap is easily destroyed~\cite{Yazyev2008,Kunstmann2011} above \mbox{$T \gtrsim 10$ K}. Therefore, both ZGNRs and CGNRs can be considered as metallic nanowires at liquid nitrogen or room temperature analyzed in our study.

We adopt the TBM with single $\pi$-orbital per site and up to third nearest-neighbor hopping
\begin{equation}\label{eq:hamilton}
\hat{H} = \sum_{\bf n} \varepsilon_{\bf n} \hat{c}_{\bf n}^\dagger \hat{c}_{\bf n} -  \sum_{{\bf n},{\bf m}} t_{\bf n}^{\bf m} \hat{c}_{\bf n}^\dagger \hat{c}_{\bf m},
\end{equation}
to describe the electronic subsystem of 20-ZGNR and (8,1)-CGNR in Fig.~\ref{fig:fig1}. The operators $\hat{c}_{\bf n}^\dag$ ($\hat{c}_{\bf n}$) create (annihilate) electron in the $\pi$-orbital located on site ${\bf n}$ of the honeycomb lattice whose lattice constant is \mbox{$a \approx 0.246$ nm} and C-C bond length is \mbox{$d_{AB} \approx 0.142$ nm}. For clean GNRs studied here the on-site potential is set to zero, $\varepsilon_{\bf n}=0$. We consider up to third nearest-neighbor~\cite{Cresti2008} hopping parameters---$t_{\bf n}^{{\bf n} + {\bf d}_{AB}}=2.7$ eV, $t_{\bf n}^{{\bf n} + {\bf d}_{AA}}=t_{\bf n}^{{\bf n} + {\bf d}_{BB}}=0.2$ eV, and $t_{\bf n}^{{\bf n} + {\bf d}_{AB^\prime}}=0.18$ eV---which describe the nearest-, next-nearest- and next-next-nearest neighbor hopping, respectively. Since the honeycomb lattice of graphene is composed of two triangular sublattices $A$ and $B$, the parameters $t_{\bf n}^{{\bf n} + {\bf d}_{AB}}$ and $t_{\bf n}^{{\bf n} + {\bf d}_{AB^\prime}}$ describe intersublattice hopping, while $t_{\bf n}^{{\bf n} + {\bf d}_{AA}}=t_{\bf n}^{{\bf n} + {\bf d}_{BB}}$ describes the intrasublattice hopping.

In the coherent transport regime, the retarded Green function (GF) matrix~\cite{Datta1995}
\begin{equation}\label{eq:gr}
{\bf G}(E) = [E{\bf I} - {\bf H} - {\bm \Sigma}_L(E) - {\bm \Sigma}_R(E)]^{-1},
\end{equation}
makes it possible to expresses the zero-bias electron transmission function between the left (L) and the right (R) electrodes as:
\begin{equation}\label{eq:telectron}
\mathcal{T}_{\rm el}(E) = {\rm Tr} \left\{ {\bm \Gamma}_R (E)  {\bf G}(E) {\bm \Gamma}_L (E)  {\bf G}^\dagger(E)  \right\}.
\end{equation}
Here ${\bf H}$ is the matrix representation of Hamiltonian in Eq.~\eqref{eq:hamilton}, ${\bf I}$ is the unit matrix in the Hilbert space of the active region, ${\bm \Sigma}_{L,R}(E)$ are the self-energies due to the ``interaction'' with the leads, and ${\bm \Gamma}_{L,R} (E) = i[{\bm \Sigma}_{L,R}(E)-{\bm \Sigma}_{L,R}^\dagger(E)]$ are the level broadening matrices determining the escape rates for electrons to enter into the attached leads. In realistic devices, active region consisting of ZGNR or CGNR of finite length with nanopores will eventually need to be connected to metallic electrodes. However, since GNR+nanopores devices we analyze are rather long $\sim 1$ $\mu$m, and screening takes place over a distance much shorter than the active region, it is justified to use semi-infinite homogeneous ZGNRs or CGNRs as leads for simplicity.

The transmission function Eq.~\eqref{eq:telectron} allows us to compute the following integrals~\cite{Esfarjani2006}
\begin{equation}\label{eq:kintegral}
K_n(\mu) = \frac{2}{h} \int\limits_{-\infty}^{\infty} dE\, \mathcal{T}_{\rm el}(E)  (E - \mu)^n \left(-\frac{\partial f(E,\mu)}{\partial E} \right),
\end{equation}
where \mbox{$f(E,\mu)=\{ 1 + \exp[(E-\mu)/k_BT] \}^{-1}$} is the Fermi-Dirac distribution function at the chemical potential $\mu$. The knowledge of $K_n(\mu)$ finally yields all electronic quantities in the expression for $ZT$: $G=e^2K_0(\mu)$; $S=K_1(\mu)/[eTK_0(\mu)]$; and $\kappa_{\rm el} = \{K_2(\mu) - [K_1(\mu)]^2/K_0(\mu)\}/T$.

The phonon subsystem is modeled using the empirical fourth-nearest-neighbor model. Its parameters were originally fitted to match the phonon dispersion measured by Raman spectroscopy~\cite{Samsonidze2003} and X-ray scattering data. Our model is reparametrized~\cite{Zimmermann2008} to include the off-diagonal terms of the force constant matrices and rotational invariance conditions which require a small correction to the in- and out-of-plane tangential force. Such empirical interatomic potential model offers excellent fit to both experiments and first-principles numerics.~\cite{Zimmermann2008}

The phonon thermal conductance, in the absence of phonon-phonon~\cite{Mingo2006} or electron-phonon~\cite{Frederiksen2007} scattering, is obtained from the phonon transmission function $\mathcal{T}_{\rm ph}(\omega)$ using the Landauer-type formula~\cite{Wang2008c}
\begin{equation}\label{eq:kappaphonon}
\kappa_{\rm ph} = \frac{\hbar^2}{2\pi k_B T^2} \int\limits_{0}^{\infty} d\omega\, \omega^2 \mathcal{T}_{\rm ph}(\omega) \frac{ e^{\hbar\omega/k_BT}}{(e^{\hbar\omega/k_BT}-1)^2}.
\end{equation}
The phonon transmission function $\mathcal{T}_{\rm ph} (\omega)$ in the elastic transport regime can be expressed in complete analogy with Eq.~\eqref{eq:telectron} for elastic electronic transport
\begin{equation}\label{eq:transphonon}
\mathcal{T}_{\rm ph}(\omega) = {\rm Tr} \left\{ {\bm \Lambda}_R (\omega)  {\bf D}(\omega) {\bm \Lambda}_L (E)  {\bf D}^\dagger(\omega)  \right\}.
\end{equation}
The phonon retarded GF is obtained in the same fashion as the electronic one in Eq.~\eqref{eq:gr} but with substitutions ${\bf H} \rightarrow {\bf K}$, $E{\bf I} \rightarrow \omega^2 {\bf M}$ and ${\bm \Sigma}_{L,R} \rightarrow {\bm \Pi}_{L,R}$
\begin{equation}\label{eq:grph}
{\bf D}(\omega)=[\omega^2 {\bf M} - {\bf K} - {\bm \Pi}_L(\omega) - {\bm \Pi}_R(\omega)]^{-1}.
\end{equation}
Here ${\bf K}$ is the force constant matrix, ${\bf M}$ is a diagonal matrix with the atomic masses, ${\bm \Pi}_{L,R}$ are the self-energies, and ${\bm \Lambda}_{L,R}(\omega)=i[{\bm \Pi}_{L,R}(\omega) - {\bm \Pi}_{L,R}^\dagger(\omega)]$. This methodology  does not take into account resistive umklapp phonon-phonon scattering which plays an important role in interpretation of experiments on room-temperature lattice thermal conductivity of large-area graphene.~\cite{Balandin2011} However, this effect,  which is easy to describe using the Boltzmann equation but is very expensive computationally within the nonequilibrium GF formalism,~\cite{Mingo2006} should not play an important role in nanoribbons  depicted in Fig.~\ref{fig:fig1} because their width is much smaller than the mean-free path \mbox{$\ell \simeq 677$ nm} due to phonon-phonon scattering in large-area graphene at room temperature.~\cite{Aksamija2011}

In conclusion, quantum transport analysis of electronic and phononic transport in ZGNR and CGNR, where electronic subsystem is described by the $\pi$-orbital tight-binding Hamiltonian with up to third nearest-neighbor hopping and phononic subsystem is described by the empirical fourth-nearest-neighbor model, suggests that these nanowires could serve as building blocks of highly efficient thermoelectric devices when perforated by an array of nanopores residing in their interior. This is due to the fact that local charge current density is peaked around their edges, as demonstrated explicitly by Fig.~\ref{fig:fig2} and confirmed experimentally,~\cite{Jia2009} so that nanopores do not impede such ``edge currents'' while  drastically reducing phonon conduction in sufficiently long ZGNRs or CGNRs. In the case of periodic array of identical nanopores, we find that  largest $ZT \simeq 3$ at $T=77$ K and  $ZT \simeq 1.5$ at $T=300$ K can be reached using (8,1)-CGNR. On the other hand, if the pore diameter takes a random value within some interval and the distance between the pores is varied, then we find a possibility to further optimize the figure of merit which can reach astonishingly large values, $ZT \simeq 11$  at $T=77$ K and $ZT \simeq 4$ at $T=300$ K, in the case of ZGNR-based device.

\begin{acknowledgments}
We thank V. Meunier for illuminating discussions.
\end{acknowledgments}

%********************references************************************************************************

%BibTeX
%Windows:
%\bibliographystyle{D:/PHYSICS/TEX/BIBTEX/prsty}
%\bibliography{D:/PHYSICS/TEX/BIBTEX/qttg}

%Linux:
%\bibliographystyle{apsrev}
%\bibliography{$HOME/TEX/BIBTEX/qttg}

\end{document}